\documentclass[11pt,a4paper]{article}
\usepackage{amsmath}
\usepackage{amsfonts}
\usepackage{amsthm}
\usepackage{amssymb}
\usepackage[T1]{fontenc}
\usepackage{graphicx}
\usepackage{multicol}
\usepackage{xspace}
\usepackage[shortlabels]{enumitem}
\usepackage{tikz}
\usepackage{bbding}
\usetikzlibrary{positioning}
\usetikzlibrary{shapes}
\usetikzlibrary{automata}
\usetikzlibrary{backgrounds}
\usetikzlibrary{fit}
\usetikzlibrary{decorations.pathmorphing}
\usetikzlibrary {arrows.meta}

\newcommand{\BN}{\ensuremath{\mathbb{N}}\xspace}

\newcommand{\CB}{\ensuremath{\mathcal{B}}\xspace}
\newcommand{\CC}{\ensuremath{\mathcal{C}}\xspace}

\newcommand{\CF}{\ensuremath{\mathcal{F}}\xspace}

\newcommand{\CK}{\ensuremath{\mathcal{K}}\xspace}
\newcommand{\CL}{\ensuremath{\mathcal{L}}\xspace}

\newcommand{\CV}{\ensuremath{\mathcal{V}}\xspace}

\newcommand\ag{ \mathsf{Ag} \xspace}

\newcommand\pow{ \mathsf{Pow} \xspace}

\newcommand\alive{ \mathsf{Alive} \xspace}
\newcommand\lalive{ \mathsf{alive} \xspace}
\newcommand\dead{ \mathsf{dead} \xspace}

\newcommand\supp{ \mathsf{supp}\xspace}
\newcommand\unds{ \mathsf{nds}\xspace}

\newtheorem{theorem}{Theorem}

\newtheorem{definition}{Definition}
\newtheorem{lemma}{Lemma}
\newtheorem{example}{Example}
\newtheorem{remark}{Remark}

\title{Simplicial Belief\thanks{This work has been funded by the Swiss National Science Foundation (SNSF). Christian Cachin and David Lehnherr are supported under SNSF grant agreement
Nr\@.~219403 (Emerging Consensus) and Thomas Studer is supported under SNSF grant agreement Nr.\@~10000440 (Epistemic Group Attitudes).
}}
\author{Christian Cachin \and
David Lehnherr  \and
Thomas Studer
}
\begin{document}
\maketitle

\begin{abstract}
Recently, much work has been carried out to study simplicial interpretations of modal logic. While notions of (distributed) knowledge have been well investigated in this context, it has been open how to model belief in simplicial models.
We introduce polychromatic simplicial complexes, which naturally impose a plausibility relation on states. From this, we can define various notions of belief.
%\keywords{Simplicial complex \and Epistemic logic \and Plausibility model \and Belief modality}
\end{abstract}

\section{Introduction}
Due to the success of applying methods from combinatorial topology to distributed computing~\cite{DBLP:books/mk/Herlihy2013}, the study of simplicial interpretations for modal logic is currently thriving~\cite{DBLP:journals/iandc/GoubaultLR21,DBLP:conf/stacs/GoubaultLR22,randrianomentsoa2023impure,10.1007/978-3-030-75267-5_1}. At its core lies the epistemic interpretation of simplicial complexes of various kinds. Let $\CV$ be a set of vertices. Each vertex corresponds to a local state of an agent, and we say that this vertex is of that agent's color. A simplicial complex $(S,\CV)$  is a pair where $S$ is a set of non-empty subsets of $\CV$ that is closed under set inclusion. Vertices that belong to the same set must be of different colors, and, in the simplest case, maximal elements of $S$ represent global states. An agent $a$ cannot distinguish two global states if its local state is included in both. Hence, simplicial complexes offer sufficient structure for an epistemic interpretation. 

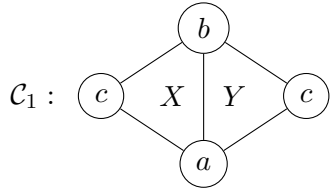
\begin{figure}[ht]
\begin{center}
\begin{tikzpicture}[scale=0.45]
\node[circle, draw] at (0,4) (b1) [label = below:$ $]{$b$};
\node[circle, draw]at (0,0) (a1)[label = above:$ $]{$a$};
\node[circle, draw]at (-3,2) (c1)[label = above:$ $]{$c$};
\node[circle, draw]at (3,2) (c2)[label = above:$ $]{$c$};

\node[circle, fill = white] at (-0.9, 2) (l1)[label = above:$ $]{$X$};
\node[circle, fill = white] at (0.9, 2) (l1)[label = above:$ $]{$Y$};
\node[circle, fill = white] at (-5, 2) (l1)[label = above:$ $]{$\CC_1:$};

\draw[-] (a1) to (b1);
\draw[-] (b1) to (c1);
\draw[-] (c1) to (a1);
\draw[-] (b1) to (c2);
\draw[-] (c2) to (a1);

\end{tikzpicture}
\caption{A simplicial complex $\CC_1$ in which the agents $a$ and $b$ cannot distinguish between the worlds $X$ and $Y$.}\label{fig:first_model}
\end{center}
\end{figure}

The simplicial complex $\CC_1$ depicted in Figure~\ref{fig:first_model} consists of two worlds $X$ and $Y$ (represented by the two solid triangles), each containing the local states of all three agents.  Agents $a$ and~$b$ have the same local state in $X$ and $Y$,  whereas $c$'s local state differs in the two global states. 
Therefore, $a$ and $b$ cannot distinguish between $X$ and $Y$, but $c$ can. That is, $a$ and $b$ always deem both states as possible, while $c$ only considers the actual state. 
This concept can be extended to groups of agents. In $\CC_1$, $a$ and $b$ together cannot distinguish between $X$ and $Y$ because they cannot do so on their own, but $a$ and $c$ can due to  $c$ being able to tell $X$ and $Y$ apart. Lastly, we say that the agents $a,b,$ and $c$ are \emph{alive} in $X$ and $Y$ because they both contain vertices with their colors.

An agent $a$ knows a formula $\phi$ in state $X$ if and only if $\phi$ is true in all states that $a$ considers as possible when in $X$.  Consequently,  if an agent is alive and knows $\phi$ in state $X$, then $\phi$ must be true in $X$.
Classical distributed knowledge~\cite{DBLP:conf/podc/HalpernM84} can be interpreted similarly. A group $G$ knows $\phi$ in a state $X$ if and only if $\phi$ is true in all states that $G$ cannot distinguish from $X$.

While (distributed) knowledge has been studied extensively in this context, it has been open, see~\cite[Section 4.3]{castaneda_et_al:DagRep.13.7.34}, how to model belief on simplicial structures. Unlike knowledge, an agent believing $\phi$ does not imply that $\phi$ is true in the actual global state. At first glance, as done in other models, this behavior can be achieved by dropping the requirement that agents must consider the actual state as possible. However, in the simplicial case, an alive agent will always consider the actual global state as possible because it contains its local state. Hence, this property cannot be dropped.

\iffalse One of the main pitfalls is that the indistinguishability relation imposed by simplicial complexes is naturally reflexive, and thus, one cannot follow the usual approach of modeling belief by dropping the reflexivity requirement of the indistinguishability relation. 
\fi

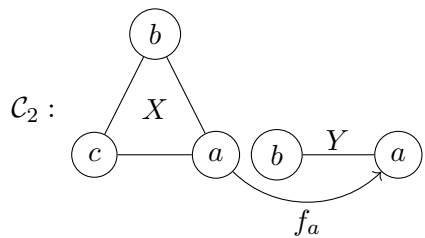
\begin{figure}[ht]
\begin{center}
\begin{tikzpicture}[scale=0.4]
\node[circle, draw] at (0,4) (b1) [label = below:$ $]{$b$};
\node[circle, draw]at (2,0) (a1)[label = above:$ $]{$a$};
\node[circle, draw]at (-2,0) (c1)[label = above:$ $]{$c$};
\node[circle, draw]at (8,0) (a2)[label = above:$ $]{$a$};
\node[circle, draw] at (4,0) (b2) [label = below:$ $]{$b$};

\node[circle, fill = white] at (5, -2.25) (l1)[label = above:$ $]{$f_a$};
\node[circle, fill = white] at (6, 0.5) (l1)[label = above:$ $]{$Y$};
\node[circle, fill = white] at (0, 1.5) (l1)[label = above:$ $]{$X$};
\node[circle, fill = white] at (-4, 1.5) (l1)[label = above:$ $]{$\CC_2:$};

\draw[-] (a1) to (b1);
\draw[-] (b1) to (c1);
\draw[-] (c1) to (a1);
\draw[-] (a2) to (b2);
\draw[->] (a1) to [out = -45, in = -135] (a2);

\end{tikzpicture}
\caption{Despite being in $X$, agent $a$ thinks that the actual world is $Y$. Thus, $a$ wrongly believes that $c$ died.}\label{fig:bfunc}
\end{center}
\end{figure}

One way of overcoming this issue is to introduce \emph{belief functions}~\cite{10.1007/978-3-030-75267-5_1}. An agent's belief function $f_a$ maps a state~$X$, in which $a$ is alive, to another state $Y$, in which $a$ is also present.  An agent $a$ believes a formula $\phi$ in $X$ if and only if it knows $\phi$ in $f_a(X)$. Hence, belief becomes knowledge if $f_a(X) = X$. 
In the model $\CC_2$ shown in Figure~\ref{fig:bfunc}, $X$ is mapped to $Y$ and $a$ believes that $c$ is not alive in the current global state, i.e., it falsely believes that $c$ no longer exists. Disadvantages of this approach are that $i)$ belief does not take the topology of the model into account and $ii)$ the principle of knowledge-yields-belief, i.e., knowing $\phi$ implies believing $\phi$, does not have a meaningful interpretation.

\iffalse
A suitable notion of belief in simplicial models should satisfy the following:
\begin{enumerate}
\item belief depends only on the topological structure of the simplicial complex;
\item the principle of knowledge-yields-belief holds.
\end{enumerate} 
\fi
In this paper, we present and interpret polychromatic simplicial complexes, i.e., complexes in which adjacent vertices may be of the same color, as simplicial models. Such models admit a notion of belief that satisfies the two conditions mentioned above. The belief studied in this work is based on the plausibility of states rather than on their possibility alone. That is, an agent believes a formula $\phi$ if and only if it is true in all states that it deems plausible enough. Since the actual global state need not be among them, the agent's beliefs might be wrong.

We start by defining a plausibility relation between the states based on the multiplicity of color within a state. If the color of an agent $a$ has a lower or equal multiplicity in a state $X$ than in a state $Y$, then $a$ considers $X$ to be at least as plausible as $Y$. If vertices are interpreted as belief states, then a possible reading of our relation is that an agent $a$ considers worlds with fewer doxastic alternatives as more plausible. 
Since this relation is a wellfounded preorder, we can use the machinery of plausibility models~\cite{baltag2006,Baltag2008} to define various notions of belief such as \emph{safe belief} and \emph{most plausible belief}.%here used to be plausible belief instead of most plausible

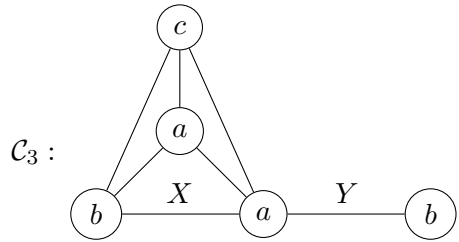
\begin{figure}[ht]
\begin{center}
\begin{tikzpicture}[scale=0.55]
\node[circle, draw] at (0,4.5) (v1) [label = below:$ $]{$c$};
\node[circle, draw]at (2,0) (v2)[label = above:$ $]{$a$};
\node[circle, draw]at (-2,0) (v3)[label = above:$ $]{$b$};
\node[circle, draw]at (0,2) (v4)[label = above:$ $]{$a$};
\node[circle, draw]at (6,0) (v5)[label = above:$ $]{$b$};

\node[circle, fill = white] at (0, 0.5) (l1)[label = above:$ $]{$X$};
\node[circle, fill = white] at (4, 0.5) (l1)[label = above:$ $]{$Y$};
\node[circle, fill = white] at (-3.5, 1.5) (l1)[label = above:$ $]{$\CC_3:$};

\draw[-] (v1) to (v2);
\draw[-] (v1) to (v3);
\draw[-] (v1) to (v4);
\draw[-] (v2) to (v3);
\draw[-] (v2) to (v4);
\draw[-] (v4) to (v3);
\draw[-] (v2) to (v5);

\end{tikzpicture}
\caption{A polychromatic simplicial complex that models the same situation as $\CC_2$ but without a belief function.}\label{fig:dead}
\end{center}
\end{figure}

Most plausible belief is a useful notion of belief. An agent $a$ most plausibly believes $\phi$ if and only if $\phi$ is true in the worlds that it considers to be the most plausible ones. This kind of belief is appropriate for reasoning about distributed computing~\cite{DBLP:books/daglib/0025983}, where agents act based on guarantees that hold true with overwhelming probability, i.e., the states in which they hold are the most plausible ones. For example, when communicating over authenticated links,  it is the most plausible case that if Alice receives a message $m$ from Bob over such a link, then $m$ was actually sent by Bob and not by an imposter. Our simplicial models can model this kind of belief while taking the topology of the model into account. The complex $\CC_3$ in Figure~\ref{fig:dead} depicts the same situation as $\CC_2$ but without a belief function. When in $X$, $a$ considers~$Y$ more plausible than $X$ because its multiplicity in $X$ is 2 and only 1 in $Y$. In this case, $a$ considers $Y$ the most plausible world. Further, since $c$ is not present in $Y$, $a$ believes that $c$ died. Thus, an agent's most plausible belief can be false. It is also straightforward to verify that this notion of belief satisfies the principle of knowledge-yields-belief. 

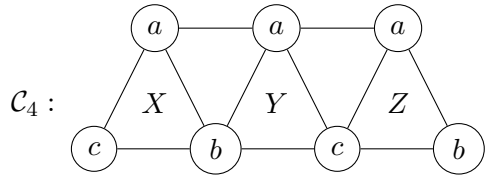
\begin{figure}[ht]
\begin{center}
\begin{tikzpicture}[scale=0.4]
\node[circle, draw] at (0,4) (a1) [label = below:$ $]{$a$};
\node[circle, draw]at (2,0) (b1)[label = above:$ $]{$b$};
\node[circle, draw]at (-2,0) (c1)[label = above:$ $]{$c$};
\node[circle, draw] at (6,0) (a2) [label = below:$ $]{$c$};
\node[circle, draw] at (4,4) (c2) [label = below:$ $]{$a$};
\node[circle, draw]at (10,0) (b2)[label = above:$ $]{$b$};
\node[circle, draw]at (8,4) (c3)[label = above:$ $]{$a$};

\draw[-] (a1) to (b1);
\draw[-] (b1) to (c1);
\draw[-] (c1) to (a1);
\draw[-] (b1) to (a2);
\draw[-] (b2) to (a2);
\draw[-] (c2) to (b1);
\draw[-] (c2) to (a2);
\draw[-] (c3) to (a2);
\draw[-] (c3) to (b2);

\draw[-] (a1) to (c2);
\draw[-] (c2) to (c3);

\node[circle, fill = white] at (8, 1.5) (l1)[label = above:$ $]{$Z$};
\node[circle, fill = white] at (4, 1.5) (l1)[label = above:$ $]{$Y$};
\node[circle, fill = white] at (0, 1.5) (l1)[label = above:$ $]{$X$};
\node[circle, fill = white] at (-4, 1.5) (l1)[label = above:$ $]{$\CC_4:$};

\end{tikzpicture}
\caption{Unicolored edges are used to represent a \emph{distance} between worlds. For example, when in $X$, $a$ considers $Y$ to be more plausible than $Z$ because the $a$-path from $X$ to $Y$ is shorter than the $a$-path from $X$ to $Z$.}\label{fig:edges}
\end{center}
\end{figure}

We close this introduction by briefly comparing our method to another, yet unexplored, approach for modeling belief on simplicial structures  mentioned in~\cite[Section 4.3.3]{castaneda_et_al:DagRep.13.7.34}. 
The idea is to impose a distance measure by allowing edges between vertices with the same color.
It is important, however, that these edges are not epistemically interpreted. The $a$-distance between two global states $X$ and $Y$ is the length of the shortest $a$-path between them. The shorter the path to a global state, the more plausible it becomes. For example, consider $\CC_4$ in Figure~\ref{fig:edges}. When in $X$, $a$ considers $Y$ more plausible than $Z$. While this interpretation takes topological information into account, it does not model false beliefs because agents always consider  the actual world as the most plausible one. 

We give an overview of the theory of simplicial models in Section~\ref{sec:simplicial_knowledge}. In Section~\ref{sec:simplicial_belief}, we introduce polychromatic models and  present different notions of belief on them. Next, we analyze in Section~\ref{sec:knowledge_gain} knowledge and belief gain in our models. In Section~\ref{sec:alternative_formulation},  we discuss other possible semantics for our models,  and, lastly, we conclude our work in Section~\ref{sec:conclusion}.

%
%\par\medskip
%\noindent
%{\bf Acknowledgements.} 
%We would like to thank the organizers and participants of the Dagstuhl meeting \emph{Epistemic and Topological Reasoning in Distributed Systems}~\cite{castaneda_et_al:DagRep.13.7.34},  especially the working group on representing epistemic attitudes via simplicial complexes.

\section{Simplicial Knowledge}\label{sec:simplicial_knowledge}

We quickly recall the standard definitions for distributed knowledge on simplicial complexes~\cite{DBLP:journals/iandc/GoubaultLR21,DBLP:conf/stacs/GoubaultLR22,10.1007/978-3-030-75267-5_1}.
In the subsequent section, we will extend them to incorporate notions of belief.

Let $\ag$ be the set of finitely many agents, and let $\mathsf{Prop}$ be a countable set of atomic propositions. We define the language of knowledge $\CL_\CK$ for $G\subseteq \ag$ and $p\in \mathsf{Prop}$ inductively by the following grammar:
\[
\phi ::= p \mid  \lnot  \phi    \mid \phi \land \phi \mid [\sim]_G \phi
\]
The remaining Boolean connectives are defined as usual. In particular, we set $\phi \lor \psi  := \lnot(\lnot \phi \land \lnot \psi)$ and $\bot:= p \land \lnot p$ for some fixed $p \in \mathsf{Prop}$. We write $\lalive(G)$ for $\neg[\sim]_G\bot$ and $\dead(G)$ for $[\sim]_G\bot$.  
\begin{definition}
Let $\CV$ be a set of vertices. $C = (S,\CV)$ with $S \subseteq \pow(\CV) \setminus \{\emptyset \}$ is called a \emph{simplicial complex} if
\[
\text{for each $X \in S$ and each $\emptyset \neq Y \subseteq X$, we have $Y\in S$.}
\]
%
% and only if 
%\begin{enumerate}[$\mathsf{S}1:$]
%\item $S$ contains all singletons, i.e., \todo{weshalb?}
%\[
%\forall v\in \CV.  \{v\} \in  S;
%\]
%\item $S$ is downward-closed, i.e.,
%\[
%\forall X\in S. Y\subseteq X\quad \text{implies}\quad Y \in S.
%\]
%\end{enumerate}
\end{definition}

\begin{remark}
The standard definition of a simplicial complex (cf.~\cite{DBLP:books/mk/Herlihy2013}) requires that the singleton of each vertex is an element of $S$. Not doing so is purely cosmetic.
\end{remark}

We call the elements of $S$ \emph{faces}. A face that is maximal under inclusion is called a \emph{facet}.  We denote the set of facets of $C$ by $\CF(C)$.
A \emph{coloring} is a mapping $\chi: \CV \rightarrow \ag$. 
A coloring is \emph{proper} if it assigns a different value to each vertex within a face.
%We call a simplicial complex that is equipped with a proper coloring \emph{chromatic}. \todo{Is this used somewhere?}
We use $\chi(U)$  for the set $\{\chi(u) \mid u\in U \}$.

~{\begin{definition}\label{def:sim_model}
Let $C =(S,\CV)$ be a simplicial complex. A \emph{simplicial model} is a quadruple $\CC = (C,\chi,W, \ell)$ where
\begin{enumerate}
\item $C$ is a simplicial complex;
\item $\chi: \CV \rightarrow \ag$ is a proper coloring;
\item $\CF(C) \subseteq W\subseteq S$ is a set of worlds;
\item $\ell: W\rightarrow \pow(\mathsf{Prop})$ is a valuation.
\end{enumerate}
\end{definition}
}

\begin{remark}
Definition~\ref{def:sim_model}  is not completely standard and is studied by Goubault et al.~\cite{GoubaultKLR2024faulty}.
Moreover, it is worth mentioning that the valuation $\ell$ in Definition~\ref{def:sim_model} assigns propositions to arbitrary faces instead of vertices only.
\end{remark}

Given a simplicial model, a group of agents $G \subseteq \ag$ cannot distinguish two worlds $X,Y\in W$, denoted by $X\sim_G Y$, if and only if $G\subseteq \chi(X\cap Y)$. We call $\sim_G$ the \emph{epistemic indistinguishability relation}. 
If we need to indicate the simplicial complex $\CC$ of which we consider the relation $X\sim_G Y$,  we use the notation $X\sim^{\CC}_G Y$.
If $G$ contains only a single agent~$a$, we write $X\sim_a Y$ and $[\sim]_a$ instead of $X\sim_{\{a\}}Y$ and $[\sim]_{\{a\}}$, respectively.

It is worth noting that $[\sim]_G$ is the usual notion of distributed knowledge of a group $G$ of agents,  which is semantically given by the intersection of the individual indistinguishability relations of the group members (cf.~\cite{DBLP:conf/podc/HalpernM84}).
\begin{definition}
%Let $C =(S,\CV)$ be a simplicial complex. 
For a simplicial model $\CC = (C,W,\chi,\ell)$, a world  $X \in W$, and a formula $\phi \in \CL_{\CK}$,  we define the relation $\CC,X \Vdash \phi$ inductively by
\begin{align*}
  & \CC,X \Vdash  p		&\text{if{f}}\qquad & p \in \ell(X)\\
    &\CC,X \Vdash  \neg \phi 			&\text{if{f}}\qquad 	& \CC,X  \not \Vdash \phi \\
  & \CC,X \Vdash \phi \land \psi  	&\text{if{f}}\qquad 	&\CC,X \Vdash \phi \text{ and } \CC,X \Vdash \psi\\
     &\CC,X \Vdash [\sim]_G \phi &\text{if{f}}\qquad & X \sim_G Y  \text{ implies }  \CC, Y \Vdash \phi \text{ for all } Y\in W.
\end{align*}
\end{definition}
We say that \emph{agent $a$ is alive in a world $X$} if $a \in \chi(X)$.
The set of worlds in which a group $G\subseteq \ag$ is alive is defined as 
\[
 \alive_\CC(G) = \{ X \in W \mid G \subseteq  \chi(X) \}.
 \]
%\[
% \alive_\CC(G) = \{ X \in W \mid {X\sim_G X} \}.
% \]
It is standard to show that $\sim_G$ is a partial equivalence relation.  We have the following lemma.

\begin{lemma}\label{lem:sim_equiv}
Let $\CC = (C,\chi,W, \ell)$ be a simplicial model. For each $G\subseteq \ag$, the relation $\sim_G$ is an equivalence relation on $\alive_\CC(G)$ and empty otherwise. 
\end{lemma}

\section{Simplicial Belief}\label{sec:simplicial_belief}

We now drop the requirement that the coloring of a simplicial model must be proper. The resulting models are called polychromatic. 
We will define a wellfounded preorder on the states of a polychromatic model, which will serve as a plausibility relation~\cite{baltag2006,Baltag2008}. This makes it possible to interpret various notions of belief on simplicial models.

It is straightforward to verify that Lemma~\ref{lem:sim_equiv} does not hold for polychromatic models because $\sim_G$ need not be transitive. 
Indeed, consider the set of vertices $\{0,1,2,3\}$ and a model that consists of the worlds 
\[
X:= \{0,1 \} , \quad  Y:= \{1,2 \} ,  \quad \text{and}  \quad Z=\{2,3 \}
\]
with a coloring $\chi$ that assigns the same agent $a$ to all vertices.
We find that $X \sim_a Y$ and $Y \sim_a Z$, but not $X \sim_a Z$.
In order to re-establish transitivity of $\sim_G$, we must require that for any three worlds $X,Y,Z \in W$ and  any group of agents $G \subseteq \ag$:
\[\label{eq:star1}
G \subseteq \chi(X \cap Y) \text{ and } G \subseteq  \chi(Y \cap Z)
\text{ implies } 
G \subseteq  \chi(X \cap Z). \tag{$\star$}
\]

\begin{definition}
A \emph{polychromatic model} is a simplicial model where:
\begin{enumerate}
\item
the coloring is not required to be proper;
\item 
condition~\eqref{eq:star1} holds.
\end{enumerate}
\end{definition}
Requiring condition~\eqref{eq:star1} is like requiring a transitive accessibility relation in certain Kripke models.

The multiplicity, see Definition~\ref{def:mult}, of a color within a face induces for each agent $a$  a wellfounded relation $\leq_a$ on worlds. We call this the (a priori) plausibility relation. 
\begin{definition}\label{def:mult}
Let $(C,\chi,W, \ell)$ be a polychromatic model. 
We define the \emph{multiplicity} of $a\in \ag$ in a world $X$ by
\[
m_a(X) = |\{ v \in X \mid \chi(v) = a\}|
\]
where $|\cdot|$ denotes the cardinality of a set.
Note that if agent $a$ is alive in a world $X$, then $m_a(X) \geq 1$.
For $X,Y \in W$ and $a\in \ag$, we write
\[
X \leq_a Y \quad \text{if{f}} \quad  m_a(X) \leq m_a(Y).
\]
\end{definition}
Next, we introduce a local plausibility relation 
\[
\unlhd_a \ :=\  \leq_a \cap \sim_a,
\]
which captures the agent's plausibility relation at a given state. Further, we write 
\[
X\geq_a Y
\quad \text{if{f}} \quad 
m_a(X)\geq m_a(Y)
\]
and we use $\unrhd_a$ and $\lhd_a$ in the obvious way.
The following lemma shows that the indistinguishability relation can be given in terms of the local plausibility relation.

\begin{lemma}\label{lem:sim_geq_equiv}
$\sim_a = \unlhd_a \cup  \unrhd_a$.
\end{lemma}
\begin{proof}
Observing that  $\leq_a$  is strongly connected and unfolding the definition yields
%Unfolding the definition yields
%\begin{align*}
%\sim_a &= \unlhd_a \cup  \unrhd_a\\
%\sim_a &= (\leq_a \cap \sim_a) \cup (\geq_a \cap \sim_a)\\
%\sim_a &= (\leq_a  \cup \geq_a )\,  \cap \sim_a.
%\end{align*}
\[
\sim_a \,=\, (\leq_a  \cup \geq_a )\,  \cap \sim_a \,=\,  (\leq_a \cap \sim_a) \cup (\geq_a \cap \sim_a) \,=\,  \unlhd_a \cup  \unrhd_a.  \qedhere
\]
%The inclusion from right to left is obvious,  the one from left to right holds since $\leq_a$ is a total order.
\end{proof}

From the relation $\unrhd_a$, we get a corresponding modal operator $[\unrhd]_a$, which is referred to in the literature as \emph{safe belief}~\cite{Baltag2008} (sometimes also as \emph{defeasible knowledge}~\cite{sep-dynamic-epistemic}).
Our language of knowledge and belief $\CL_{\CK\CB}$ extends  $\CL_\CK$ by the modal operator~$[\unrhd]_a$ for each agent $a\in \ag$. It  is inductively defined by as follows:
\[
\phi ::= p \mid  \lnot  \phi    \mid \phi \land \phi \mid [\sim]_G \phi \mid [\unrhd]_a \phi
\]
where $p \in \mathsf{Prop}$ and $G \subseteq  \ag$.
As usual, the dual of safe belief is defined as $\langle \unrhd\rangle_a \varphi \equiv \neg [\unrhd]_a \neg \varphi$. 
%Further,  we let $\lalive(G)$ be the formula $\bigwedge_{a \in G}\lnot [\sim]_a \bot$.
%
%
% In the literature, the induced modal operator $[\unlhd]_a$ is referred to safe-belief. The dual of safe belief is defined as $\langle \unlhd\rangle_a \varphi \equiv \neg [\unlhd]_a \neg \varphi$. Lastly, we write $ X\geq_a Y$ if and only if   $m_X(a)\geq m_Y(a)$ and we use $\geq_a, <_a,>_a, \unrhd_a$ and $\lhd_a$ in the obvious way. Lemma~\ref{lem:sim_geq_equiv} shows that knowledge can be expressed with $\unlhd_a$.
%
%\begin{lemma}\label{lem:sim_geq_equiv}
%$\sim_a = \unlhd_a \cup  \unrhd_a$.
%\end{lemma}
%
%We now enrich the language $\CL_\CK$ with the operator $[\unlhd]_a$ and obtain $\CL_{\CK\CB}$ which allows us to reason about the beliefs of a single agent (see Definition \ref{def:belief}). 

\begin{definition}
For a polychromatic model $\CC=(C,\chi,W, \ell)$, a world  $X \in W$, and a formula $\phi \in \CL_{\CK\CB}$, we define the relation $\CC,X \Vdash \phi$ inductively by
\begin{align*}
  & \CC,X \Vdash  p		&\text{if{f}}\qquad & p \in \ell(X)\\
    &\CC,X \Vdash  \neg \phi 			&\text{if{f}}\qquad 	& \CC,X  \not \Vdash \phi \\
  & \CC,X \Vdash \phi \land \psi  	&\text{if{f}}\qquad 	&\CC,X \Vdash \phi \text{ and } \CC,X \Vdash \psi\\
     &\CC,X \Vdash [\sim]_G \phi &\text{if{f}}\qquad & X \sim_G Y \quad \text{implies} \quad \CC, Y \Vdash \phi \text{, for all } Y\in W\\
      &\CC,X \Vdash [\unrhd]_a \phi &\text{if{f}}\qquad &  X\unrhd_a Y \quad \text{implies} \quad \CC, Y \Vdash \phi \text{, for all } Y\in W.
\end{align*}
\end{definition}

%We write $\CC \Vdash \phi$, if  $\CC,X \Vdash \phi$ for all $X \in w$.
%A formula $\phi$ is valid, denoted by $ \Vdash \phi$, if $\CC \Vdash \phi$ for all models $\CC$.

The  $[\unrhd]_a$-modality satisfies the $\mathsf{S4.2}$ principles  for alive agents.

\begin{lemma}\label{l:s42:1}
The following formulas are valid:
\begin{enumerate}
\item $[\unrhd]_a(\phi \to \psi) \to ([\unrhd]_a \phi \to  [\unrhd]_a \psi)$;
\item $\lalive(a) \to ([\unrhd]_a \phi \to \phi)$;
\item $[\unrhd]_a \phi \to  [\unrhd]_a  [\unrhd]_a \phi $;
\item $\langle \unrhd\rangle_a [\unrhd]_a \phi \to  [\unrhd]_a \langle \unrhd\rangle_a  \phi $.
\end{enumerate}
\end{lemma}
\begin{proof}
We will only show the last claim.  Assume 
\begin{equation*}\label{eq:S42:1}
\CC,X \Vdash  \langle\unrhd\rangle_a [\unrhd]_a \phi.
\end{equation*}
There exists $Z$ with $X \unrhd_a Z$ and
\begin{equation}\label{eq:S42:2}
\CC,Z \Vdash  [\unrhd]_a \phi.
\end{equation}
From $X \unrhd_a Z$ we get that $a$ is alive in $Z$ and thus $Z \unrhd_a Z$.  Therefore,  by \eqref{eq:S42:2}
\begin{equation}\label{eq:S42:3}
\CC,Z \Vdash  \phi.
\end{equation}
Let $Y$ be arbitrary with $X  \unrhd_a Y$.
By \eqref{eq:star1}, we find $Y \sim_a Z$.  Thus we get by Lemma~\ref{lem:sim_geq_equiv}, that 
$Z \unrhd_a Y$ or  $Y \unrhd_a Z$.
In the first case,  we use \eqref{eq:S42:2} to obtain $\CC,Y \Vdash \phi$.  Further we get $Y  \unrhd_a Y$ from $X  \unrhd_a Y$, and thus 
\begin{equation}\label{eq:S42:4}
\CC,Y \Vdash  \langle\unrhd\rangle_a \phi.
\end{equation}
In the second case,  \eqref{eq:S42:4} follows immediately from \eqref{eq:S42:3}.
Since $Y$ was arbitrary  with $X  \unrhd_a Y$,  \eqref{eq:S42:4}  implies
$\CC,X \Vdash   [\unrhd]_a \langle \unrhd\rangle_a  \phi$. 
\end{proof}

As usual with plausibility models, not only can we define safe belief but also other notions of belief. We start by defining the set of most plausible worlds.

\begin{definition}
Let $\CC = (C,\chi,W, \ell)$  be a polychromatic model. For $X\in W$, we define
\[
\mathsf{Min}_{\unlhd_a}(X) =
 \{Y\in W \mid Y\sim_a X \text{ and } \nexists Z \in W. Z \lhd_a Y\}.
\]
\end{definition}
%Since $\leq_a$ is wellfounded, we find that $\mathsf{Min}_{\unlhd_a}(X) \neq \emptyset$ if agent $a$ is alive in the world~$X$.

Observe that, in general,  since $\leq_a$ is wellfounded, $\mathsf{Min}_{\unlhd_a}(X) \neq \emptyset$ if agent $a$ is alive in the world~$X$.

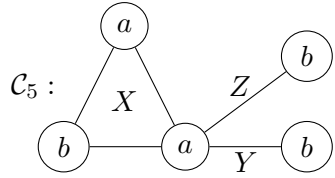
\begin{figure}[ht]
\begin{center}
\begin{tikzpicture}[scale=0.4]
\node[circle, draw] at (0,4) (v1) [label = below:$ $]{$a$};
\node[circle, draw]at (2,0) (v2)[label = above:$ $]{$a$};
\node[circle, draw]at (-2,0) (v3)[label = above:$ $]{$b$};
\node[circle, draw]at (6,0) (v4)[label = above:$ $]{$b$};
\node[circle, draw]at (6,3) (v5)[label = above:$ $]{$b$};

%\node[circle, draw] at (11,4) (v5) [label = below:$ $]{$a$};
%\node[circle, draw]at (13,0) (v6)[label = above:$ $]{$a$};
%\node[circle, draw]at (9,0) (v7)[label = above:$ $]{$b$};

\node[circle, fill = white] at (-3,2) (l1)[label = above:$ $]{$\CC_5:$};
%\node[circle, fill = white] at (8,2) (l1)[label = above:$ $]{$C^\psi:$};
\node[circle, fill = white] at (4,-0.5) (l1)[label = above:$ $]{$Y$};
\node[circle, fill = white] at (0,1.5) (l1)[label = above:$ $]{$X$};
%\node[circle, fill = white] at (11,1.5) (l1)[label = above:$ $]{$X$};
\node[circle, fill = white] at (3.8,2) (l1)[label = above:$ $]{$Z$};

\draw[-] (v1) to (v2);
\draw[-] (v2) to (v3);
\draw[-] (v3) to (v1);
\draw[-] (v2) to (v4);
\draw[-] (v2) to (v5);

%\draw[-] (v5) to (v6);
%\draw[-] (v6) to (v7);
%\draw[-] (v7) to (v5);

\end{tikzpicture}
\caption{A polychromatic model with two minimal worlds.}\label{fig:minimal}
\end{center}
\end{figure}

The model $\CC_5$ given in Figure~\ref{fig:minimal} shows a situation with two minimal worlds.  Namely,  we have $Y, Z \in \mathsf{Min}_{\unlhd_a}(X)$.
Further, the following relations hold: ${Y \lhd_a X}$,  $Z \lhd_a X$,  $Z \unlhd_a Y$, and $Y \unlhd_a Z$.

The following lemma about minimal worlds will be useful.
\begin{lemma}\label{l:obs:1}
Let $\CC = (C,\chi,W, \ell)$  be a polychromatic model,  $X\in W$,  and $Y \unlhd_a X$.
For each $Z \in \mathsf{Min}_{\unlhd_a}(X)$,  we have $Z \unlhd_a Y$.
\end{lemma}
\begin{proof}
From $Y \unlhd_a X$ we get
$X \sim_a  Y$. % and $Y \leq_a X$. 
Let $Z \in \mathsf{Min}_{\unlhd_a}(X)$.  By definition, this means
\begin{equation}\label{eq:obs:1}
\nexists V \in W. V \lhd_a Z
\end{equation}
and $X \sim_a  Z$.  
Since $\sim_a$ is a partial equivalence relation, we get  $Z \sim_a  Y$. 
Suppose  towards a contradiction that $Z \unlhd_a Y$ does not hold. 
Since $Z \sim_a  Y$ holds, we must have $Z\nleq_a Y$ and thus $Y <_a Z$.
Therefore,  $Y \lhd_a Z$.   
This contradicts~\eqref{eq:obs:1}, and
we conclude $Z \unlhd_a Y$. 
\end{proof}
In the above lemma, we can let $Y$ be $X$.  Then we obtain the following instance:
\begin{equation}\label{eq:obs:2}
Z \in \mathsf{Min}_{\unlhd_a}(X) \quad\text{implies}\quad Z \unlhd_a X.
\end{equation}
We  now include a new modality  $ \CB_a$ for each agent $a$  in our language $\CL_{\CK\CB}$.
We extend the  truth definition for $\CL_{\CK\CB}$ as follows.
\begin{definition}\label{def:belief}
 For a polychromatic model $\CC=(C,\chi,W, \ell)$, a world  $X \in W$, and a formula $\phi \in \CL_{\CK\CB}$, we define 
\[
\CC, X \Vdash \CB_a \varphi 
\text{ if{f} } 
\quad Y\in \mathsf{Min}_{\unlhd_a}(X) \text{ implies } \CC, Y \Vdash \varphi \quad \text{for all } Y\in W.
\]
\end{definition}

The modality $ \CB_a$  models agent $a$'s (most plausible) belief.
It is well-known that  $ \CB_a$  can be expressed in terms of the $[\unrhd]_a$-modality~\cite{Baltag2008,stalnaker2006}. In our setting, we have the following theorem.

~\begin{theorem} Let\/ $\CC = (C,\chi,W, \ell)$ be a polychromatic model, $a$ an agent, and $X\in W$ such that $a$ is alive in $X$.
We find that  
\begin{center}
$\CC, X \Vdash \CB_a\varphi$ \quad if and only if \quad $\CC, X \Vdash \langle \unrhd  \rangle_a [\unrhd ]_a \varphi$.
\end{center} 
\end{theorem}
\begin{proof}
For the direction from right to left,  we assume $\CC, X \Vdash \langle \unrhd  \rangle_a [\unrhd ]_a \varphi$.
Thus, there exists $Y$ with $X \unrhd_a Y$ and $\CC, Y  \Vdash [\unrhd ]_a \varphi$.
Now consider an arbitrary $~{Z \in \mathsf{Min}_{\unlhd_a}(X)}$.
By Lemma~\ref{l:obs:1} we find $Y \unrhd_a Z$ and,  therefore,  $\CC, Z  \Vdash  \varphi$.
This yields $\CC, X \Vdash \CB_a\varphi$.

For the direction from left to right, we have $X \sim_a X$ since agent $a$ is alive.
Thus  $\mathsf{Min}_{\unlhd_a}(X)$ is non-empty and we let
 $Y \in \mathsf{Min}_{\unlhd_a}(X)$.
By~\eqref{eq:obs:2},  we obtain $Y \unlhd_a X$.
It remains to show $\CC, Y \Vdash  [\unrhd ]_a \varphi$.
Let $Z \in W$ be such that  $Z \unlhd_a Y$.
Then $Z \sim_a X$ by transitivity of $\sim_a$.
Now we find $Z \in \mathsf{Min}_{\unlhd_a}(X)$, for otherwise, 
we would find $V \in W$ with $V \lhd_a Z$,  which yields  $V \lhd_a Y$ and thus contradicts
$Y \in \mathsf{Min}_{\unlhd_a}(X)$.
From $Z \in \mathsf{Min}_{\unlhd_a}(X)$ and the assumption  $\CC, X \Vdash \CB_a\varphi$, we get
$\CC, Z \Vdash \varphi$, which concludes the proof.
\end{proof}

\begin{remark}
A consequence of this theorem is that the properties of the $\CB_a$-modality follow from properties of\/ $[\unrhd ]_a$ such as the ones given in Lemma~\ref{l:s42:1}. For instance, we find that
\[
\CB_a \phi \land  \CB_a \psi \to  \CB_a (\phi \land \psi).
\] 
is valid. 
\end{remark}
%% See 
%% https://mathoverflow.net/questions/160355/a-question-on-the-modal-logic-s4-2

Our model satisfies the knowledge yields belief principle. In particular, we have the following lemma.
\begin{lemma}
Let $\CC = (C,\chi,W, \ell)$  be a polychromatic model and $X\in W$. For any agent $a$ and any formula $\varphi$, we have 
\[
\CC, X \Vdash [\sim]_a \varphi \rightarrow[\unrhd ]_a  \varphi \quad\text{and} \quad
 \CC, X \Vdash [\unrhd ]_a  \varphi \rightarrow \CB_a \varphi.
\]
\end{lemma}
\begin{proof}
For the first claim, assume 
\begin{equation}\label{eq:l4:1}
\CC, X \Vdash [\sim]_a \varphi .
\end{equation}
Let $X \unrhd_a Y$,  i.e., $X \sim_a  Y$ and $X \geq _a Y$.
By \eqref{eq:l4:1} we immediately get $\CC, Y \Vdash \varphi$ and hence,   $\CC, X \Vdash [\unrhd ]_a \varphi$.

For the second claim,  assume
\begin{equation}\label{eq:l4:2}
\CC, X \Vdash[\unrhd ]_a  \varphi .
\end{equation}
Let $Y \in \mathsf{Min}_{\unlhd_a}(X)$ be arbitrary.
Using~\eqref{eq:obs:2}, we obtain 
$Y \unlhd_a X$.
Now $\CC, Y \Vdash \varphi$ follows immediately from \eqref{eq:l4:2},  which 
yields
$\CC, X \Vdash  \CB_a \varphi$.
\end{proof}
%Assume $\CC, X \Vdash [\sim]_a \varphi$. By definition, $Y\unlhd_a X$ implies $X\sim_a Y$. Hence, by assumption,  $C,Y \Vdash \varphi$ for any $Y\in \mathsf{Min}_{\unlhd_a}(X)$. Therefore, $~{C, X \Vdash \CB_a \varphi}$. 
%\end{proof}

As usual with preference-based semantics,  our models are non-monotone. That is, agents may drop their beliefs when learning new information. 
\begin{example}\label{ex:nonm:1}
Consider the set $\CV = \{1,2,3,4\}$ and let $C = (S,\CV)$ and $~{C^\psi = (S^\psi, \CV)}$  be simplicial complexes given by 
\[
S = (\pow(\{1,2,3\})\cup \pow(\{3,4\}))\setminus \{\emptyset\} \text{ and }S^\psi = \pow(\{1,2,3\})\setminus \{\emptyset\}.
\] 
Further, let $X = \{1,2,3\}$ and $Y = \{3,4\}$. We define the models
\[
\CC = (C,\chi,\{X,Y\}, \ell \}) \text{ and } \CC^\psi = (C^\psi, \chi, \{X\}, \ell^\psi).\]
The coloring $\chi$ is defined as follows
\begin{enumerate}
\item $\chi(1)  = \chi(3) = a$;
\item $\chi(2)  = \chi(4) = b$.
\end{enumerate} 
Figure~\ref{fig:non-monotonicity} shows the colored complexes.
We choose the valuation $\ell$ such that for some propositional formulas $\psi,\phi \in \CL_{\CK\CB}$:
\[
\CC, X  \Vdash \neg\phi \land \psi \quad\text{and}\quad
\CC, Y \Vdash \phi \land \neg \psi
\]
and we define $\ell^\psi(X) = \ell (X)$. The model $\CC^\psi$ represents the  situation  after the agents in $\CC$ learn that $\psi$ is true. That is, it is the same as $\CC$ but without the worlds where $\psi$ is false. We observe that 
\[
\CC, X \Vdash \CB_a \phi \quad \text{and}\quad \CC^\psi, X \not \Vdash \CB_a \phi.
\]
Hence, $a$ only believes $\phi$ in $X$ before it learns $\psi$.
This is because removing worlds from $\CC$ can result in a new world becoming a most plausible world. 
%
%This cannot happen in the usual Kripke models for belief where $\sim_a^{\CC^\psi} \subseteq \sim_a^\CC $ for all $a\in \ag$.

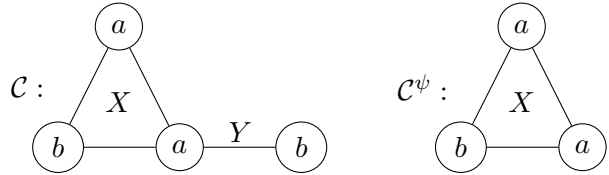
\begin{figure}[ht]
\begin{center}
\begin{tikzpicture}[scale=0.4]
\node[circle, draw] at (0,4) (v1) [label = below:$ $]{$a$};
\node[circle, draw]at (2,0) (v2)[label = above:$ $]{$a$};
\node[circle, draw]at (-2,0) (v3)[label = above:$ $]{$b$};
\node[circle, draw]at (6,0) (v4)[label = above:$ $]{$b$};

\node[circle, draw] at (13.25,4) (v5) [label = below:$ $]{$a$};
\node[circle, draw]at (15.25,0) (v6)[label = above:$ $]{$a$};
\node[circle, draw]at (11.25,0) (v7)[label = above:$ $]{$b$};

\node[circle, fill = white] at (-3,2) (l1)[label = above:$ $]{$\CC:$};
\node[circle, fill = white] at (10,2) (l1)[label = above:$ $]{$\CC^\psi:$};
\node[circle, fill = white] at (4,0.5) (l1)[label = above:$ $]{$Y$};
\node[circle, fill = white] at (0,1.5) (l1)[label = above:$ $]{$X$};
\node[circle, fill = white] at (13.25,1.5) (l1)[label = above:$ $]{$X$};

\draw[-] (v1) to (v2);
\draw[-] (v2) to (v3);
\draw[-] (v3) to (v1);
\draw[-] (v2) to (v4);

\draw[-] (v5) to (v6);
\draw[-] (v6) to (v7);
\draw[-] (v7) to (v5);

\end{tikzpicture}
\caption{The polychromatic complex $\CC^\psi$ represents the state of affairs after the agents in $\CC$ learn that $\psi$ is true.}\label{fig:non-monotonicity}
\end{center}
\end{figure}
\end{example}
A model is called \emph{proper} if different worlds can be  distinguished by at least one agent. Properly colored simplicial models are proper (cf.~\cite{GoubaultKLR2024faulty}). Formally, this is expressed as
\begin{equation}\label{eq:proper:1}
\lalive(G) \land \dead(G^c) \land \varphi \rightarrow [\sim]_G (\dead(G^c) \rightarrow \varphi)
\end{equation}
being valid, where $G^c$ stands for the complement of $G$.  Lemma~\ref{lem:not-proper} shows that this is no longer true for polychromatic models.

\begin{lemma}\label{lem:not-proper}
Polychromatic models are not proper, i.e. \eqref{eq:proper:1} is not valid.
\end{lemma}

\begin{proof}
Consider the following counter-example.
Let $C =(S,\CV)$ be the simplicial complex given by
\[
\CV=\{1,2,3\} \quad\text{and}\quad S= \pow(\CV) \setminus \{\emptyset\}.
\]
Let $a,b$ be agents.  We set 
\[
\chi(1):=\chi(2):=a \quad\text{and}\quad \chi(3):=b .
\]
The set of possible worlds $W$ contains only the two worlds
$X:=\{1,2,3\}$ and $Y:=\{1,3\}$. The valuation is such that $\ell(X):=\{p\}$,  and $\ell(Y):=\emptyset$.

Let $\CC := (C,\chi,W, \ell)$. For $G=\{a,b\}$, we find that
\[
\CC,X \Vdash \lalive(G) \land \dead(G^c) \land p.
\]
However,  we have $X \sim_G Y$ and $\CC,Y \Vdash \dead(G^c) \land \lnot p$.  Hence
\[
\CC,X \not\Vdash [\sim]_G (\dead(G^c) \rightarrow p).
\]
Therefore,  \eqref{eq:proper:1} is not valid on polychromatic models.
\end{proof}

\section{Knowledge Gain}\label{sec:knowledge_gain}

An important result for epistemic simplicial models is that agents cannot gain new knowledge along morphisms of simplicial models.  This property is crucial in distributed computing to show that certain morphisms do not exist,  which implies that certain computation tasks are not solvable.

We adapt the notion of morphism from~\cite{DBLP:journals/iandc/GoubaultLR21} to the setting where models contain a set of worlds.  The fact that our models are polychromatic does not matter for the definition of morphism. 
As usual,  given a function $f: U \rightarrow V$ and a set $W \subseteq U$,  we let 
\[
f(W):=\{f(x) \ |\ x \in W\}.
\]
%As for notation, let $U$ and $V$ be two sets of vertices and let $f: U \rightarrow V$. For $W \subseteq U$ and 
%\[
%S = \bigcup_{w\in W}f(w),
%\]
%we write $f(\pow(W))$ for $\pow(S)$. 
\begin{definition}
Let $C =(S,\CV)$ and\/ $C' =(S',\CV')$ be two simplicial complexes.
A \emph{simplicial map} from $C$ to $C'$ is a function $f:\CV \to \CV'$ such that if $X\in S$ then $f(X)\in S'$.
\end{definition}

\begin{definition}\label{def:morph:1}
Let $C =(S,\CV)$ and\/ $C' =(S',\CV')$ be two simplicial complexes and
let $\CC = (C,\chi,W, \ell)$ and $\CC' = (C',\chi',W', \ell')$ be two polychromatic models.
A \emph{morphism} from $\CC$ to\/ $\CC'$ is a function $f$ such that
\begin{enumerate}
\item $f$ is a simplicial map from $C$ to $C'$;
\item $\chi'(f(v)) = \chi(v)$ for all $v \in \CV$;
\item $f(X) \in W'$ for all $X \in W$;
\item $\ell'(f(X)) = \ell(X)$ for all $X \in W$.
\end{enumerate}
\end{definition}

Morphisms respect the indistinguishability relation.  We have the following lemma.
\begin{lemma}\label{l:morph:1}
Let $C =(S,\CV)$ and\/ $C' =(S',\CV')$ be two simplicial complexes and
let $\CC = (C,\chi,W, \ell)$ and $\CC' = (C',\chi',W', \ell')$ be two polychromatic models and $f$ be a morphism  from $\CC$ to\/ $\CC'$.
For $X,Y \in W$ we have
\[
X \sim^{\CC}_G Y \quad\text{implies}\quad f(X) \sim^{\CC'}_G f(Y).
\]
\end{lemma}
\begin{proof}
Assume
$G\subseteq \chi(X\cap Y)$ and let $a$ be an element of $G$.
There exists $v\in \CV$ with $v \in X$,  $v \in Y$,  and $\chi(v)=a$.
We find that $f(v) \in f(X)$,  $f(v) \in f(Y)$,  and $\chi'(f(v))=a$.
Hence,  $a \in \chi'(f(X)\cap f(Y))$.
\end{proof}

The positive formulas are the formulas of $\CL_\CK$ where the operator $[\sim]_G$ occurs only in positive positions. Formally,  we use the following definition.
\begin{definition}\label{def:positive}
We consider the following grammar:
\begin{align*}
\phi &::= p \mid  \lnot  \phi    \mid \phi \land \phi\\
\psi &::=   \phi    \mid \psi \land \psi  \mid \psi \lor \psi \mid [\sim]_G  \psi 
\end{align*}
where $p\in \mathsf{Prop}$ and $G\subseteq \ag$.
Formulas given by $\psi$ are called \emph{positive formulas}.
\end{definition}

The result about no knowledge gain is standard~\cite{DBLP:journals/iandc/GoubaultLR21}.
Note that we are in a setting where agents may crash.  However, since we assign atomic variables to the worlds (and not the local states of the agents),  we can employ the usual formulation of positive formulas in the following theorem.
Also, the fact that we have polychromatic models does not matter.

\begin{theorem}
Let  $\CC = (C,\chi,W, \ell)$ and $\CC' = (C',\chi',W', \ell')$ be two polychromatic models and $f$ be a morphism from  $\CC$ to $\CC'$.
Further,  let $\psi$ be a positive formula. 
We find that 
\[
\CC', f(X) \Vdash \psi \quad\text{implies}\quad \CC, X \Vdash \psi
\]
for all $X \in W$.
\end{theorem}
\begin{proof}
First we show that for a formula given by $\phi$ according to Definition~\ref{def:positive},  we have 
\begin{equation}\label{eq:guarded:1}
\CC', f(X) \Vdash \phi \quad\text{if{f}}\quad \CC, X \Vdash \phi
\end{equation}
for all $X \in W$.
We proceed by induction on the structure of $\phi$ and distinguish:
\begin{enumerate}
\item $\phi$ is an atomic proposition $p$.  Since $f$ is a morphism,  we have $~{\ell'(f(X)) = \ell(X)}$.  Thus $\CC, f(X) \Vdash p$ if{f} $\CC, X \Vdash p$.
\item $\phi$ is a negation or a conjunction.  The claim follows immediately by I.H.
\end{enumerate}

Now we proceed by induction on the structure of $\psi$ (acc.~to Def.~\ref{def:positive})
and assume $\CC', f(X) \Vdash \psi$. We distinguish the following cases:
\begin{enumerate}
\item $\psi$ is a formula $\phi$  (according to Definition~\ref{def:positive}). The claim follows from~\eqref{eq:guarded:1}.
\item $\psi$ is a conjunction or a disjunction. The claim follows immediately by the induction hypothesis.
\item $\psi$ is  of the form $[\sim]_G  \psi'$.  
From 
$~{\CC', f(X) \Vdash [\sim]_G  \psi'}$ we obtain that 
$ f(X)\sim^{\CC'}_G Y$ implies  $~{\CC', Y \Vdash \psi'}$ for all $Y \in W'$.

Let $Z\in W$ be such that $X \sim^{\CC}_G Z$.  
Since $f$ is a morphism,  we find by Lemma~\ref{l:morph:1} that 
\[
f(X) \sim^{\CC'}_G f(Z) \quad\text{and}\quad f(Z) \in W'.
\]
Thus,   $\CC', f(Z) \Vdash \psi'$. 
By I.H., we find  $\CC,  Z \Vdash \psi'$. 
Since $Z\in W$ was arbitrary with $X \sim^{\CC}_G Z$, we conclude  $\CC, X \Vdash [\sim]_G  \psi'$.
\qedhere
\end{enumerate}
\end{proof}

The previous theorem only holds for knowledge but not for belief. That is, the operator $ [\unrhd ]_a$ cannot be included in the class of positive formulas. We have the following lemma.
\begin{lemma}
There are two polychromatic models  $\CC$ and\/ $\CC'$, a morphism $f$  from $\CC$ to\/ 
$\CC'$ and a world $X$ of $\CC$ such that for some agent $a$ and $p\in \mathsf{Prop}$,
\[
\CC',f(X)\Vdash  [\unrhd ]_a p  \quad\text{but}\quad \CC,X\not\Vdash  [\unrhd ]_a p.
\]
\end{lemma}
\begin{proof}
Consider the set $\CV=\{1,2,3\}$.  We let $a$ be an agent and set $$\chi(1):=\chi(2):=\chi(3):=a.$$
We let $S:=\{ \{1,2\},\ \{2, 3\}, \ \{1\},\ \{2\},\ \{3\}   \}$ and $S':=\{ \{1,2\},\ \{1\},\ \{2\}  \}$.
Hence  $C =(S,\CV)$ and  $C' =(S',\CV)$ are simplicial complexes.
We let 
$X:=\{1,2\}$,   $Y:=\{2,3\}$,  and $Z:=\{2\}$. 
Further, we set
\[
W:= \{ X,Y \}, \qquad \ell(X):=\emptyset \quad\text{and}\quad  \ell(Y):= \{p\}
\]
and
\[
W':= \{ X, Z \},  \quad \ell'(X):=\emptyset \quad\text{and}\quad  \ell'(Z):=\{p\}.
\]
We define the polychromatic models
\[
\CC := (C,\chi,W, \ell)\quad\text{and}\quad \CC' := (C',\chi,W', \ell')
\]
and let $f$ be such that $f(1)=1$,  $f(2)=2$, and $f(3)=2$.
Obviously,  $f$ is a morphism from $\CC$ to $\CC'$ and we have $f(Y)=Z$.
Finally, we note that
\[
\CC',f(Y)\Vdash  [\unrhd ]_a p  \quad\text{but}\quad \CC,Y\not\Vdash  [\unrhd ]_a p.   \qedhere
\]
\end{proof}
%
%
%%%%%%%%%%%%%%
%\par\bigskip
%%%%%%%%%%%%
%\begin{proof}
%Let $C =(S,\CV)$ be the simplicial complex given by
%\[
%\CV=\{1,2\} \quad\text{and}\quad S=\{ \{1,2\},\ \{1\},\ \{2\} \}.
%\]
%Let $a$ be an agent.  We set $\chi(1):=\chi(2):=a$.
%We let $X:=\{1,2\}$ and $Y:=\{1\}$. 
%Further we set
%\[
%W:= \{ X \} \quad\text{and}\quad  \ell(X):=\emptyset
%\]
%and
%\[
%W':= \{ X, Y \},  \quad \ell'(X):=\emptyset \quad\text{and}\quad  \ell'(Y):=\{p\}.
%\]
%We define the polychromatic models
%\[
%\CC := (C,\chi,W, \ell)\quad\text{and}\quad \CC' := (C,\chi,W', \ell')
%\]
%and let $f$ be the identity function.  Obviously,  $f$ is a morphism from $\CC$ to $\CC'$ and we have $f(X)=X$.
%Finally we note that
%\[
%\CC',f(X)\Vdash  [\unrhd ]_a p  \quad\text{but}\quad \CC,X\not\Vdash  [\unrhd ]_a p.
%\]
%\end{proof}
Besides the statement that belief gain is possible,  this lemma could also be  interpreted in such a way that condition~\eqref{eq:star1} is not strong enough or that we need a different notion of morphism,  see also the discussion about simplicial sets in the next section.

%This lemma can be seen as a converse to Example~\ref{ex:nonm:1}. There, we removed a world, which led to fewer beliefs. In the above proof, we added a world and obtained additional beliefs.

\section{Alternative Formalizations}\label{sec:alternative_formulation}

The important feature of our approach to model belief with simplicial models is that agents are assigned a multiplicity within a world.
Polychromatic models achieve this by allowing several vertices to represent the same agent.
We have chosen this approach because it has the advantages of being very close to the original epistemic interpretation of simplicial complexes and having a simple geometric  interpretation. Compared to~\cite{GoubaultKLR2024faulty}, only the constraint of unique colors within a world has been relaxed.  However, to guarantee that the indistinguishability relation is a partial equivalence relation,  we have to impose condition~\eqref{eq:star1}  on polychromatic models.

In this section, we present two alternative approaches to belief in simplicial models that are based on the idea of having repeated vertices instead of having several vertices with the same color.  In these settings, the vertices of a world all have different colors, and thus condition~\eqref{eq:star1} is not needed.
 
We will only sketch the basic ideas and leave an elaborate presentation for future work.
The first approach consists of using a version of simplicial complex that is based on multisets instead of sets.  One could use the following definitions.

Let $\CV$ be a set of vertices.
A multiset over $\CV$ is a function $M: \CV \to  \BN$.
The empty multi-set~$\emptyset$ denotes the constant 0-function.
A multiset $N$ is a subset of a multiset $M$,  in symbols as usual $N \subseteq M$,  if  we have $N(v) \leq M(v)$ for all $v \in \CV$.
The support of a  multiset $M$ is given by the set of its elements.,  i.e.~$\supp(M):=\{ v \ |\ M(v) >0\}$.

~\begin{definition}
$C = (S,\CV)$ with 
$S$ being a  set of multisets over\/ $\CV$ is called a \emph{multi-simplicial complex} if
\begin{enumerate}
\item $S$ does not contain the empty set;
\item for each $X \in S$ and each $\emptyset \neq Y \subseteq X$, we have $Y\in S$.
\end{enumerate}
\end{definition}

A \emph{multi-simplicial model} is then defined like a simplicial model where a multi-simplicial complex is used instead of a simplicial complex.
Since a world in a multi-simplicial model is a multiset,  we can define the 
 \emph{multiplicity} of an agent $a$ in a world $X$ by
%\[
%m_a(X) := \max\{ X(v) \mid \chi(v) = a \text{ and } v \in \CV \}.
%\]
\[
m_a(X) := \begin{cases}  X(v) & \text{if there exists $v$ with $\chi(v) = a$ and  $X(v)>0$,}\\
0 & \text{otherwise.}\end{cases}
\]
Note that by the definition of multi-simplicial model,  where the coloring must be proper, 
\begin{equation}\label{eq:multi:1}
\text{there is at most one $v \in \CV$ with $\chi(v) = a$ and $X(v)>0$.}
\end{equation}
Hence,  $m_a(X)$ is well-defined.

We can employ the usual indistinguishability relation if we use the support of the worlds, meaning we consider the worlds as sets instead of multisets.
For a group of agents $G$,  we define 
\[
X\sim_G Y \quad\text{if{f}}\quad G\subseteq \chi(\supp(X) \cap \supp(Y))
\]
Using~\eqref{eq:multi:1}, we immediately find that the so-defined relation $\sim_G$ is transitive.  Thus, we do not need to additionally impose a condition like~\eqref{eq:star1}. The disadvantage of multi-simplicial models is that they do not have a topological interpretation.

\par\medskip

The second alternative approach,  which does have a natural topological interpretation, is to work with simplicial sets,  see,  e.g.~\cite{10.1216/RMJ-2012-42-2-353} for an introduction. 
For our very informal discussion here,  it is sufficient to think of simplicial sets as simplices that may contain repeated vertices.
Hence,  the set of worlds of a simplicial set model may contain simplices as well as simplices with repeated vertices, which are called \emph{degenerate} simplices.

It is a standard result~\cite[Prop.~4.8]{10.1216/RMJ-2012-42-2-353}
that if $Y$ is a degenerate simplex,  then there is  a unique non-degenerate simplex~$X$ such that $Y = s_{i_1}\cdots s_{i_k}X$ for some collection of degeneracy maps $s_{i_1},\ldots,s_{i_k}$.
% Eilenberg- Zilber lemma [GZ67, pp 26-27]
% https://math.jhu.edu/~eriehl/ssets.pdf
Here,  a degeneracy map creates one copy of a vertex. 
Given a degenerate simplex~$Y$, we denote the unique non-degenerate simplex $X$ of the above result by $\unds(Y)$.  If $Y$ is a non-degenerate simplex, we set $\unds(Y):=Y$.

Using this setting, we can define the indistinguishability relation for simplicial set models by
\[
X\sim_G Y \quad\text{if{f}}\quad G\subseteq \chi(\unds(X) \cap \unds(Y)).
\]
As before,  $\sim_G$ is transitive, and condition~\eqref{eq:star1} is not needed.
Moreover, we can define the multiplicity of a vertex $v$ in a world $Y$ as 1 plus the  number of copies of $v$ that the degeneracy maps introduce in the construction of $Y$ from $ \unds(Y)$.  

Note that we have to be careful when we move from polychromatic models to simplicial set models.  One of the reasons is that we have different notions of morphisms for the two models.
The notion of morphism for polychromatic models (Definition~\ref{def:morph:1}) is based on simplicial maps that preserve the color.  However, the multiplicity of an agent need not be preserved  along morphisms for polychromatic models.
The situation is different for simplicial set models.  If we take the notion of morphism from simplicial sets and require that the color is preserved,  then the multiplicity of an agent will also be preserved.

So far,  simplicial sets have not been considered models for epistemic logic.  However, we want to mention that semi-simplicial sets, which lie between simplicial complexes and simplicial sets,  have been used to model novel notions of group knowledge~\cite{DBLP:journals/tcs/CachinLS25,goubault2023semisimplicial}.

\section{Conclusion}\label{sec:conclusion}

We presented the first interpretation of belief on simplicial models that depends only on the topological structure without requiring additional machinery like belief functions.
Our approach consists of dropping the requirement that the coloring must be proper and using the multiplicity of color within a face as an inverted plausibility measure.

The study of polychromatic models is still in its infancy. This paper presents first definitions, initial results, and key differences to simplical models such as the observations about non-proper models and belief gain.

Although we included a notion of  group knowledge in our logic,  we only considered individual belief.
Obviously,   notions of group belief  provide an interesting topic for future research.
A first definition could be
\[
X \leq_G Y \text{ if and only if }
\min\{m_a(X) \mid a\in G\} \leq \min\{m_a(Y) \mid a\in G\}
\]
and then define the interpretation of the modalities 
$[\unrhd]_G$ and  $\CB_G$ as in the individual case.
Some principles of this notion of group belief are immediate,  e.g. that group belief does not imply belief of subgroups (or individual belief).
However, a detailed analysis of this approach should be performed and its relationship to existing approaches, e.g.~\cite{Gaudou2015,DBLP:journals/corr/abs-2408-10637}, should be studied.

\begin{figure}[h]
\begin{center}
\begin{tikzpicture}[scale=0.3]
\node[circle, draw] at (0,0) (a0) [label = below:$ $]{$a$};
\node[circle, draw] at (12,0) (a1) [label = left:$ $]{$a$};
\node[circle, draw] at (6,0) (b0) [label = left:$ $]{$b$};
\node[circle, draw] at (6,-5.2) (c0) [label = left:$ $]{$c$};
\node[circle, draw] at (17.2,2.6) (c1) [label = left:$ $]{$c$};
\node[circle, draw] at (17.2,-2.6) (b1) [label = left:$ $]{$b$};
\node[circle, draw] at (-5.2,2.6) (c2) [label = left:$ $]{$c$};
\node[circle, draw] at (-5.2,-2.6) (b2) [label = left:$ $]{$b$};
\draw[-] (a0) to node[midway, below, rotate = -60]{$ $} (b0);
\draw[-] (c0) to node[midway, below, rotate = 0]{$ $} (a1);
\draw[-] (a1) to node[midway, below, rotate = 60]{$ $} (c1);
\draw[-] (c1) to node[midway, below, rotate = -60]{$ $} (b1);
\draw[-] (b1) to node[midway, below, rotate = 0]{$ $} (a1);
\draw[-] (b0) to node[midway, below, rotate = 60]{$ $} (c0);
\draw[-] (b0) to node[midway, below, rotate = -60]{$ $} (a1);
\draw[-] (a0) to node[midway, below, rotate = 0]{$ $} (c0);
\draw[-] (a0) to node[midway, below, rotate = 60]{$ $} (b2);
\draw[-] (c2) to node[midway, below, rotate = -60]{$ $} (b2);
\draw[-] (c2) to node[midway, below, rotate = 0]{$ $} (a0);
\draw[-] (a1) to[out=-225, in=45]  node[midway, below, rotate = 0]{$Z$} (a0);

\node[-] at (-3,0) (b2) [label = left:$ $]{$X_0$};
\node[-] at (15,0) (b2) [label = left:$ $]{$Y_0$};
\node[-] at (7,-2.6) (b2) [label = left:$ $]{$Y_1$};
\node[-] at (5,-2.6) (b2) [label = left:$ $]{$X_1$};
\end{tikzpicture}
\caption{Polychromatic complexes in which simplices are either properly colored or unicolored might be used for developing a simplicial neighborhood semantics.
}\label{fig:neighbourhood}
\end{center}
\end{figure}
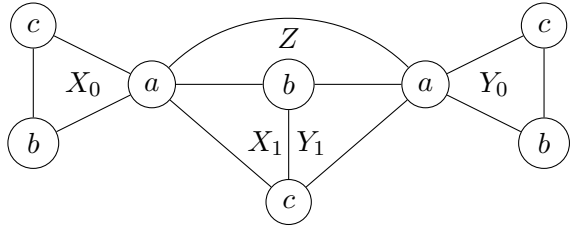

Another possible research direction for polychromatic models is to interpret them as a simplicial neighborhood semantics~\cite{castaneda_et_al:DagRep.13.7.34}.
One approach would be to consider models in which simplices are either colored properly or unicolored. 
In Figure~\ref{fig:neighbourhood},  for instance,  the simplices $X_0,X_1,Y_0,Y_1$ are properly colored (each vertex has a different color) whereas $Z$ is unicolored (every vertex has the same color).

Unicolored simplices can be used to define a neighborhood structure. 
The simplex $X_0$ has two $a$-neighborhoods $\{X_0,X_1\}$ and $\{Y_0,Y_1\}$ because agent $a$ can reach $Y_0$ via the unicolored edge $Z$. 
It is straightforward to see that such neighborhoods will never intersect in $a$.
The interpretation of knowledge is then standard: agent $a$ knows $\varphi$ if there is an $a$-neighborhood in which $\varphi$ is true in all worlds.

Last but not least,  let us mention that recently,  a variant of topological semantics was introduced for polyhedra~\cite{ADAM-DAY_BEZHANISHVILI_GABELAIA_MARRA_2024,polyhedra,BEZHANISHVILI2018373}.  It will be interesting to investigate the similarities and differences between simplicial semantics and polyhedral semantics for modal logic.

\section*{Acknowledgements}
We thank the anonymous reviewers for helpful suggestions and
feedback.
Special thanks goes to the organizers and participants of the Dagstuhl meeting \emph{Epistemic and Topological Reasoning in Distributed Systems}~\cite{castaneda_et_al:DagRep.13.7.34}, especially the working group on representing epistemic attitudes via simplicial complexes.
We would also like to thank the anonymous reviewers of our AiML short paper~\cite{aiml2024} for many helpful comments.

%This work has been funded by the Swiss National Science Foundation (SNSF). Christian Cachin and David Lehnherr are supported under SNSF grant agreement
%Nr\@.~219403 (Emerging Consensus) and Thomas Studer is supported under SNSF grant agreement Nr.\@~10000440 (Epistemic Group Attitudes).

%\subsubsection{\discintname}
%The authors have no competing interests to declare that are
%relevant to the content of this article. 

%\bibliography{references}
%\bibliographystyle{splncs04}

\end{document}